# Regge resonances in low-energy electron elastic cross sections for Ge, Sn and Pb atoms: Manifestations of stable excited anions

A.Z. Msezane[1], Z. Felfli[1] and D. Sokolovski[2]

[1]Department of Physics and Centre for Theoretical Studies of Physical Systems,
Clark Atlanta University, Atlanta, Georgia 30314, USA
[2]School of Mathematics and Physics, Queen's University of Belfast,
Belfast, BT7 1NN, UK

**Abstract**

Low-energy $E < 2$ eV electron elastic collisions with Ge, Sn and Pb atoms yield stable excited Ge⁻, Sn⁻ and Pb⁻ anions. The recent Regge-pole methodology is used with Thomas-Fermi type potential incorporating the crucial core-polarization interaction to calculate elastic total and Mulholland partial cross sections. For excited Ge⁻ and Sn⁻ anions the extracted binding energies from the unique characteristic sharp Regge resonances manifesting stable excited states formed during the collisions agree excellently with experimental values; for Pb⁻ the prediction requires experimental verification. The calculated differential cross sections also yield the binding energies.

**PACS Numbers:** 34.80.Bm; 34.50.Cx

Here we consider low-energy $E < 2$ eV elastic collisions between electrons and neutral atoms and use for the first time ever the recent Regge-pole methodology to investigate the possibility of forming and observing stable excited negative ions, as Regge resonances. If the total energy of the negative ion thus formed is greater than that of the neutral atom, then the negative ion is embedded in the continuum of states of the free electron and the atom [1], forming a temporary negative ion as a resonance. Electron-electron and core-polarization interactions are two important fundamental physical mechanisms responsible for the existence and stability of most negative ions. For heavy and complex atoms, generally characterized by intricate and subtle interactions among the many diverse electron configurations, obtaining reliable theoretical electron affinities (EAs) is very difficult, if not impossible because of the computational complexity encountered by theoretical investigations of the structure and dynamics of such atoms using structure-based type calculations. Binding energies, numerically equal to the EAs, if obtained through these methods are often riddled with uncertainties; hence the use of the Regge-pole methodology which requires no *a priori* knowledge of the EAs whatsoever.

The many meanings of resonance in the literature lead to much confusion [2]. Here we use the rigorous definition of resonances as singularities of the S-matrix [3, 4] to explore low-energy electron attachment to Ge, Sn and Pb atoms forming excited negative ionic states, manifesting as Regge resonances, through the Mulholland formula [5], implemented within the complex angular momentum (CAM), L representation of scattering. The recently developed Regge-pole analysis [6] which embodies the essential electron correlations together with a Thomas-Fermi (TF) type potential, incorporating the vital core-polarization interaction, are used to calculate the electron elastic total and the Mulholland partial cross sections. The differential cross sections (DCSs) are calculated using a partial wave expansion. The power and utility of the Regge-pole methodology in predicting reliably the binding energies (BEs) for stable bound states of tenuously bound (BE < 0.1 eV), weakly bound (BE < 1 eV) and complicated open d- and f-sub-shell negative ions, have now been firmly established [7-10].

Furthermore, the recent investigations [11, 12] have demonstrated that the electron DCSs in scattering angle $\theta$ and in impact energy are capable of yielding the BEs of the negative ions formed during the collisions as resonances.

Infrared laser spectroscopy (ILS) [13] has been employed to investigate the bound excited np³ ²D terms of the Si⁻, Ge⁻ and Sn⁻ negative ions and the BEs of the two J=3/2 and J=5/2 fine-structure levels were obtained, while in [14] and [15] the term averaged BEs for Ge⁻ and Sn⁻ and for Si⁻, respectively were determined. The experimentalists [13, 16] also obtained from laser photodetachment threshold (LPT) spectroscopy the BEs of the ground state of the Si⁻, Ge⁻ and Sn⁻ anions, while Miller *et al.* [17] used laser photodetachment electron spectroscopy (LPES) to measure the BEs for the Ge⁻ and Sn⁻ ions. No BE values for either the ground state or the excited state of the Pb⁻ ion was reported because of the difficulty of producing a Pb⁻ ion beam [13]. Generally, theoretical calculations are extremely complex, laborious and computationally intensive, yielding uncertain results. So, to our knowledge, the BEs are nonexistent for the excited Pb⁻ anion.

Here the recent Regge-pole methodology [6] is first benchmarked on the accurately measured BEs of the excited bound states of the Ge⁻ and Sn⁻ anions [13] through the BEs of these anions that are extracted from the Regge-pole calculated TCSs. Then we extract the BE of the excited bound state of the Pb⁻ negative ion from the resonances in the calculated TCS using the Regge pole methodology. The obtained BEs for the excited states of the Ge⁻ and Sn⁻ anions agree excellently with those of the measurements [13]; the BE for the excited Pb⁻ anion requires experimental verification. It is further demonstrated that these BEs can also be extracted directly from the DCSs at the scattering angles $\theta$ = 0°, 90° and 180°, giving experimentalists a new simple and direct approach to measuring the BEs of excited bound anion states. We note that the measurement obtained only the BEs of the excited Ge⁻ and Sn⁻ anions, while here we extract the BEs from the TCSs and DCSs, which are presented here.

For the investigation of the formation of excited states in near-threshold electron atom collisions the Mulholland formula [5] is employed in the form [6, 18] (atomic units are used throughout):

$$\sigma_{tot}(E) = 4\pi k^{-2} \int_0^\infty \text{Re}[1-S(\lambda)]\lambda d\lambda$$
$$-8\pi^2 k^{-2} \sum_n \text{Im} \frac{\lambda_n \rho_n}{1+\exp(-2\pi i\lambda_n)} + I(E) \quad (1)$$

where S is the S-matrix, k = √(2mE), with m being the mass, $\rho_n$ the residue of the S-matrix at the nth pole, $\lambda_n$ and I(E) contains the contributions from the integrals along the imaginary λ-axis; its contribution has been demonstrated to be negligible [10]. We will consider the case for which Im $\lambda_n$<<1 so that for constructive addition, $\text{Re}\lambda_n \approx 1/2, 3/2, 5/2....$, yielding $\ell = \text{Re} L \cong 0,1,2...$ The importance of Eq. (1) is that a resonance is likely to affect the elastic TCS when its Regge pole position is close to a real integer [6]. We also need the DCS given by

$$d\sigma(E,\theta)/d\Omega = |f(E,\theta)|^2, \quad (2)$$

where the scattering amplitude $f(E,\theta)$ can be expressed as a partial wave sum

$$f(E,\theta) = (1/2ik)\sum_{\ell=0}^\infty (2\ell+1)P_\ell(\cos\theta)[S_\ell(E)-1] \quad (3)$$

with $k, \ell, \theta, P_\ell(\cos\theta)$ and $S_\ell(E)$ being the wave vector, the angular momentum quantum number, the scattering angle, a Legendre polynomial of degree $\ell$ and the scattering matrix element, respectively. For central field scattering

$$S_\ell(E) = \exp[i\delta(E, \ell)], \tag{4}$$

where $\delta(E, \ell)$ are the phase shifts.

For the calculation of both the TCSs and the DCSs the TF-type potential [19] takes the well investigated [20] form

$$U(r) = \frac{-Z}{r(1 + aZ^{1/3}r)(1 + bZ^{2/3}r^2)}, \tag{5}$$

where $Z$ is the nuclear charge and $a$ and $b$ are adjustable parameters. For small r, the potential describes the Coulomb attraction between an electron and a nucleus, $U(r) \sim -Z/(r)$, while at large distances it mimics the polarization potential, $U(r) \sim -1/(abr^4)$ and accounts properly for the vital core-polarization interaction at very low energies. The effective potential

$$V(r) = U(r) + L(L+1)/(2r^2), \tag{6}$$

is considered here as a continuous function of the variables $r$ and L. The potential, Eq. (5) has been used successfully with the appropriate values of $a$ and $b$. When the TCS as a function of "$b$" has a resonance [10] corresponding to the formation of a stable bound negative ion, this resonance is longest lived for a given value of the energy which corresponds to the electron affinity of the system. This was found to be the case for all the systems we have investigated thus far. This fixes the optimal value of "$b$" for Eq. (5). In the case of Ge, Sn and Pb, we have found two such values of $b$ which still satisfy the TF equation within a certain error margin. We identify the smallest value with an excited state and the largest with the ground state of the corresponding atom. Note that the energies are measured relative to the ground-state of the neutral atom. The values used for "$b$" in this paper are 0.0410, 0.0470 and 0.0348 for the excited Ge, Sn and Pb atoms, respectively, while the value of "$a$" was kept fixed at 0.2 for the three atoms. For the ground state of the Pb atom the optimal value of "$b$" is 0.0493 and was used for the lower curve of Fig. 1(d).

For the numerical evaluation of the TCSs and the Mulholland partial cross sections, we solve the Schrödinger equation for complex values of L and real, positive values of E

$$\psi'' + 2\left(E - \frac{L(L+1)}{2r^2} - U(r)\right)\psi = 0, \tag{7}$$

with the boundary conditions:

$$\psi(0) = 0,$$
$$\psi(r) \sim e^{+i\sqrt{2E}r}, \; r \to \infty. \tag{8}$$

We note that Eq. (8) defines a bound state when $k \equiv \sqrt{(2E)}$ is purely imaginary positive. In solving Eq. (7) two independent approaches are adopted. The first integrates numerically the radial Schrödinger equation for real integer $\ell = \text{Re } L$ values of L to sufficiently large r values. The S-matrix is then obtained and the TCSs and the DCSs are evaluated as the traditional sum over partial waves, with the index of summation being $\ell$. The second part calculates the S-matrix, S(L, k) poles positions and residues of Eq. (7) following a method similar to that of Burke and Tate [21]. In the method the two linearly independent solutions, $f_L$ and $g_L$, of the Schrödinger equation are evaluated as Bessel functions of complex order and the S-matrix, which is defined by the asymptotic boundary condition of the solution of the Schrödinger equation, is thus evaluated. Further details of the calculation may be found in [21].

Experimental studies of the near-threshold electron attachment mechanism in neutral atoms, resulting in the formation of excited states of atomic negative ions that are bound with respect to the atomic ground state are very limited [13], and hardly exist theoretically. This has severely limited the knowledge of the existence and understanding of excited negative ions. From a theoretical perspective, excited negative ions are generally weakly bound systems with BE < 2 eV, making structure-based type calculations very difficult because of the many diverse and complex configurations involved. Since the recent Regge-pole approach to low-energy electron attachment requires no *a priori* knowledge of their EAs, the Regge-pole methodology promises to enhance their exploration and advance their understanding.

Figure 1(a) presents the near-threshold electron elastic scattering total and Mulholland partial cross sections for the excited Ge atom, showing the Mulholland contribution to the TCS. The collision is characterized by two distinct resonances. The first peak corresponds to a shape resonance at 0.054 eV, while the second sharp peak represents the stable bound excited state of the Ge⁻ anion with the BE of 0.342 eV, formed during the collision. From the Regge trajectories [10] (not shown) we found that Re L=1 and Im L=0.052 for the shape resonance while Re L=3 and Im L=$1.8 \times 10^{-4}$ for the stable bound excited state. The values of the Im L demonstrate the usual identification that bound states are several orders of magnitude smaller than those for the shape resonances [10]. The electron attaches to a Re L=3 in forming the stable bound excited Ge⁻ anion. The extracted BE for the stable bound excited Ge⁻ anion compares excellently with the term averaged value from the measurement of Scheer *et al.* [13] and very well with the term averaged value obtained by Hotop and Lineberger [14].

The electron elastic total and Mulholland partial cross sections for the excited Sn atom, showing the Mulholland contribution to the TCS is displayed in Fig. 1(b). Here also the collision is characterized by two distinct resonances, the first and second are at 0.054 eV and 0.340 eV, respectively. Just as in the case of the e⁻ - Ge scattering, they correspond, respectively to a shape resonance and the stable bound excited Sn⁻ anion, formed during the collision. From the Regge trajectories (not shown) we found that Re L=1 and Im L=0.042 for the shape resonance while Re L=3 and Im L= $1.3 \times 10^{-4}$ for the stable bound excited state. As usual the values of the Im L demonstrate the identification that bound states are several orders of magnitude smaller than those for the shape resonances. In this case the electron attaches to a Re L=3 in forming the stable bound excited Sn⁻ anion. The extracted BE for the stable bound excited Sn⁻ anion compares excellently with the term averaged value from the measurement [13] and very well with the term averaged value obtained in [14].

The Regge pole calculated electron elastic total and Mulholland partial cross sections for the excited Pb atom, showing the Mulholland contribution to the TCS, are presented in Fig. 1(c). The characteristic two resonance peaks seen above for the e⁻ - Ge and e⁻ - Sn scattering represent, respectively the shape resonance at 0.022 eV and the stable bound excited Pb⁻ anion at 0.103 eV. The calculated Regge trajectories yielded Re L= 1 and Im L≈ 0.020 for the shape resonance and Re L=3 and Im L=$7.1 \times 10^{-6}$ for the stable bound excited Pb⁻ anion, demonstrating that the electron attaches to a Re L=3 in forming the stable bound excited Pb⁻ anion. No published experimental or theoretical BEs for the Pb⁻ anion are available, to our knowledge, to compare with, not to mention the TCSs.

In Fig. 1(d) is contrasted the electron elastic scattering TCSs for ground state (lower curve) and excited state (upper curve) Pb atoms; the two curves are quite different, hence the importance of the comparison. The former curve exhibits the characteristic behavior found in weakly bound systems [10], manifesting a Ramsauer-Townsend minimum at 0.102 eV, a shape resonance at 0.346 eV and the stable bound ground Pb⁻ anion at 0.563 eV formed during the collision as a Regge resonance. We note that this value of the BE is higher than the experimental one [13]. The difference between the measured and the nonrelativistic EA for Pb has been explained in terms of the claimed significant spin-orbit interaction in Pb [23]. However, the difference 0.563 eV - 0.340 eV =0.223 eV (difference between the present value and that of [13]) is much smaller than the 0.77 eV attributed to the spin-orbit interaction by the recent theoretical calculation [23]. The full explanation will be provided in a forth coming paper [22] dealing with electron scattering from ground states of Ge, Sn and Pb in which TCSs as well as DCSs in both energy and scattering angle are presented.

The DCSs provide stringent test of theoretical calculations when the results are compared with those of reliable measurements; therefore they will be useful in this context, particularly with respect to the BEs of the excited Ge⁻ and Sn⁻ anions since their accurate measurements are available [13]. Figures 2(a), (b) and (c) present the DCSs (*a. u.*) in energy at θ = 0°, 90° and 180° for the electron elastic scattering from the excited Ge, Sn and Pb atoms, respectively. Most significant about these curves is that the DCSs are characterized by sharp distinct peaks at θ = 0° and θ = 90°, corresponding to the values of the BEs, *viz.* 0.342, 0.340 and 0.103 eV, respectively of the stable bound excited states of the Ge⁻, Sn⁻ and Pb⁻ anions, formed during the collision as resonances. The DCSs at θ = 180° also yield the BEs, but the peaks have now become anti-resonances. Preceding the stable bound excited states of the Ge⁻, Sn⁻ and Pb⁻ anions are the broader peaks that correspond to the shape resonances as described above under the TCSs. The DCSs critical minima [24, 25] are also identifiable at 1.26 eV, 1.50 eV and 1.19 eV, respectively; the results in the figures have meanings only for E< 2 eV. Although the DCSs appear quite similar in the figures, there are some subtle differences among them. For example, in Fig. 2(c) the dramatically sharp peak at θ = 0° and 90° is much higher than the corresponding ones in Figs. 2(a) and 2(b) and the curve at θ = 180° shows a small maximum and a deep minimum at the BE of the stable bound excited state of the Pb⁻ anion, in contrast to those in Figs. 2(a) and 2(b).

Low-energy electron elastic collisions with Ge, Sn and Pb atoms resulting in electron attachment as Regge resonances have been investigated for the first time ever using the recent Regge pole methodology. The Im L has been used to differentiate between the shape resonances and the resultant stable bound excited states of the anions. Benchmarking the BEs extracted from the Regge-pole calculated electron elastic TCSs for excited Ge and Sn atoms on the accurately measured BEs for the excited states of the Ge⁻ and Sn⁻ anions [13], the methodology has been used to predict the BE of the stable bound excited Pb⁻ anion. The results for the excited Pb⁻ anion call for experimental verification.

Electron elastic scattering DCSs in energy at θ = 0°, 90° and 180° for the excited Ge, Sn and Pb atoms have also been calculated using a partial wave expansion and the BEs of the stable bound excited states of the Ge⁻, Sn⁻ and Pb⁻ anions have been determined. The distinct sharp characteristic resonances in the calculated DCSs at these angles are identified with the BEs of the resultant excited negative ions formed during the collision as resonances. Thus the DCSs provide a new powerful approach to both experimental and theoretical investigations of electron elastic scattering from both ground [12] and excited atoms resulting in the formation of weakly bound ionic states.

**Acknowledgments**

Research was supported by U.S. DOE, Division of Chemical Sciences, Office of Basic Energy Sciences, Office of Energy Research and the CAU CFNM, NSF-CREST Program. DS is supported through a EPSRC (UK) Grant. The computing facilities at the Queen's University of Belfast, UK and of DOE Office of Science, NERSC are greatly appreciated.

**Figure Captions**

Figure 1(a): Total and Mulholland partial elastic cross sections, in atomic units, for e⁻ - Ge scattering versus E (eV), showing the Mulholland contributions. The first peak at 0.054 eV corresponds to a shape resonance, while the second one at 0.342 eV represents the stable excited bound state of the Ge⁻ anion formed during the collision as a resonance.

Figure 1(b): Same as in Fig. 1 (a), except that the results belong to the electron elastic scattering from Sn. Here the shape resonance is at 0.054 eV while the BE at 0.340 eV corresponds to the stable excited bound state of the Sn⁻ anion formed during the collision as a resonance.

Figure 1(c): Same as in Fig. 1 (a), except that the results are for the electron elastic scattering from Pb. Here the shape resonance is at 0.022eV while the BE at 0.103 eV corresponds to the stable excited bound state of the Pb⁻ anion formed during the collision as a resonance.

Figure 1(d): The total cross sections, in atomic units, for e⁻ - Pb scattering versus E (eV), are contrasted, lower curve is for the ground state and the upper curve is for the excited state. The cross sections have validity only for E < 2 eV. The lower curve has the characteristic RT minimum at 0.102 eV, a shape resonance at 0.346 eV and a stable bound ground state of the Pb⁻ anion at 0.563 eV (see comment in the text about this value).

Figure 2(a): The DCSs for electron elastic scattering from Ge versus E (eV) at the scattering angles $\theta = 0°, 90°$ and $180°$. At all the scattering angles the shape resonance and the BE of the stable excited bound state of the Ge⁻ anion formed during the collision as a resonance can be determined readily. Note the presence of the DCS critical minimum at about 1.26 eV; the results are meaningful only for E< 2 eV.

Figure 2(b): The same as in Fig. 2(a), except that the data are for electron elastic scattering from Sn and the stable excited bound state corresponds to the Sn⁻ anion and the critical minimum is at about 1.50 eV.

Figure 2(c): The same as in Fig. 2(a), except that the data are for electron elastic scattering from Pb and the stable excited bound state corresponds to the Pb⁻ anion and the critical minimum is at about 1.19 eV.

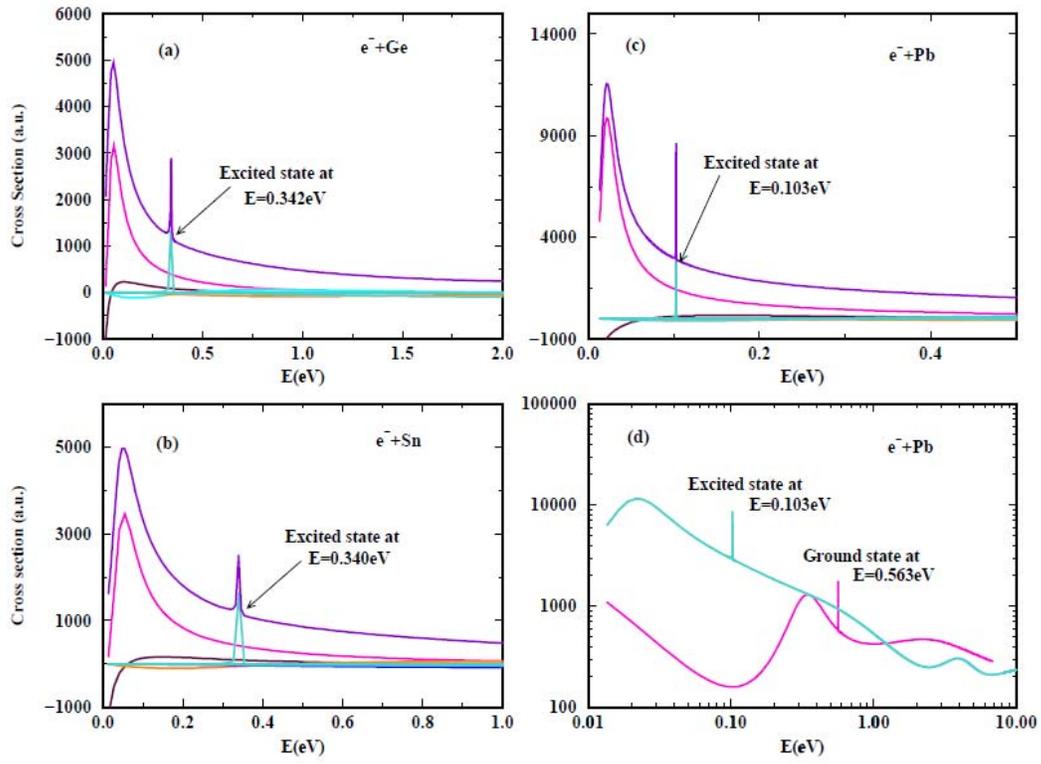

**Fig. 1**

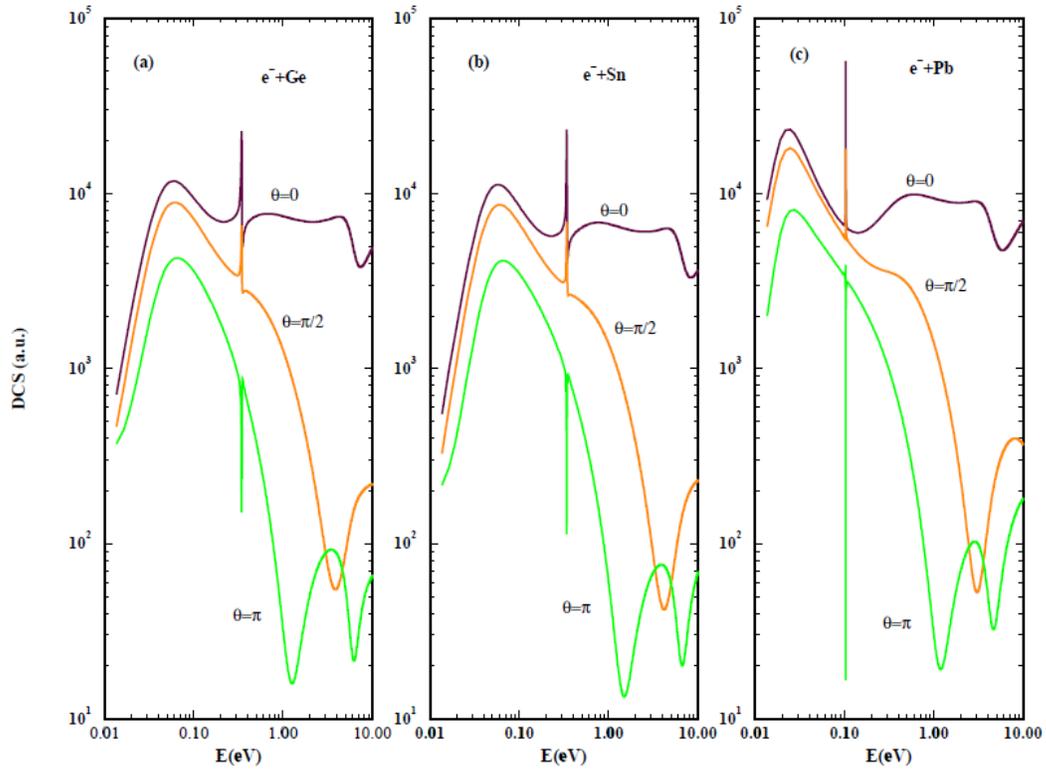

**Fig. 2**